\def\fr#1#2{\hbox{${#1\over #2}$}}
\def\prg#1{\medskip{\bf #1}}        \def\lra{\leftrightarrow}
\def\Ra{\Rightarrow}                
\def\ni{\noindent}                  \def\pd{\partial}
\def\wtilde{\widetilde}             \def\mb#1{\hbox{\boldmath$#1$}}

\def\m{\mu}             \def\n{\nu}              
\def\G{\Gamma}          \def\g{\gamma}           \def\d{\delta}
\def\S{\Sigma}          \def\s{\sigma}           \def\t{\tau}
\def\a{\alpha}          \def\b{\beta}            \def\th{\theta}
      \def\ve{\varepsilon}     \def\p{\pi}
\def\r{\rho}            \def\D{\Delta}           \def\L{{\Lambda}}
\def\l{{\lambda}}       \def\om{\omega}          \def\Om{\Omega}

\def\cL{{\cal L}}               \def\cP{{\cal P}}
      \def\cO{{\cal O}}        
\def\cA{{\cal A}}
\def\tcL{\wtilde\cL}          

\def\bG{\mb{\Gamma}}    \def\bL{\mb{L}}
\def\bull{{\vrule height.8ex width0ex}\raise.15ex\hbox{\vrule
                                        height.8ex width.8ex}}
\def\Chr#1#2{\hbox{${#1\brace #2}$} }
\def\tgr{GR$_{\parallel}$}
\def\pt{{\hbox{\sc\scriptsize pt}}}
\documentclass[11pt]{article}
\usepackage{latexsym,graphicx,equations}  

\def\nn{\nonumber}
\def\be{\begin{equation}}             \def\ee{\end{equation}}
\def\ba#1{\begin{array}{#1}}           \def\ea{\end{array}}
\def\bea{\begin{eqnarray} }           \def\eea{\end{eqnarray} }
\def\lab#1{\label{eq:#1}}             \def\eq#1{(\ref{eq:#1})}
\def\bsubeq{\begin{subequations}}     \def\esubeq{\end{subequations}}
\def\bitem{\begin{itemize}}           \def\eitem{\end{itemize}}
\renewcommand{\theequation}{\thesection.\arabic{equation}}
\def\mypreprint{
\renewcommand{\theequation}{\thesection.\arabic{equation}}
\textwidth 16.4cm            \textheight 23.2cm
\evensidemargin=-.3cm        \oddsidemargin=-.3cm
\topmargin=-1cm
}

\mypreprint
\begin{document}
\title{\Large\bf Three lectures on Poincar\'e gauge theory\thanks{
       Based on lectures presented at II Summer School
       in Modern Mathematical Physics, Kopaonik, Yugoslavia,
       1-12 September, 2002.} }
\author{M. Blagojevi\'c\thanks{Email address: mb@phy.bg.ac.yu}
       \vspace{3pt}\\
       {\it\small Institute of Physics, 11001 Belgrade,
            P\/.O\/.\,Box 57, Yugoslavia}\\
       {\it\small 
            Primorska Institute for Nat. Sci. {\rm\&} Tech.\/,
            6000 Koper, P\/.O\/.\,Box 327, Slovenia}}
\date{}
\maketitle\vspace{-1em}
\begin{abstract}
In these lectures we review the basic structure of Poincar\'e gauge
theory of gravity, with emphasis on its fundamental principles and
geometric interpretation. A specific limit of this theory, defined by
the teleparallel geometry of spacetime, is discussed as a viable
alternative for the description of macroscopic gravitational phenomena.
\end{abstract}

\section*{Introduction}

Despite its successes in describing {\it macroscopic\/} gravitational
phenomena, Einstein's general re\-la\-ti\-vi\-ty (GR) still lacks the
status of a fundamental {\it microscopic\/} theory, because of the
problem of quantization and the existence of singular solutions under
very general assumptions. Among various attempts to overcome these
difficulties, gauge theories of gravity are especially attractive, as
the concept of gauge symmetry has been very successful in the
foundation of other fundamental interactions. The importance of
Poincar\'e symmetry in particle physics leads one to consider
Po\-in\-ca\-r\'e gauge theory as a natural framework for description of
the gravitational phenomena.

The principle of equivalence implies that Einstein's GR is invariant
under local Poincar\'e transformations. Instead of thinking of local
Poincar\'e symmetry as derived from the principle of equivalence, the
whole idea can be reversed, in accordance with the usual philosophy of
gauge theories. When gravitational field is absent, it has become clear
from a host of experiments that the underlying spacetime symmetry of
fundamental interactions is given by the {\it global\/} (rigid)
Poincar\'e group. If we now want to make a physical theory invariant
under {\it local\/} Poincar\'e transformations, with parameters which
depend on spacetime points, it is necessary to introduce new, {\it
compensating fields\/}; these fields are found to represent the
gravitational interaction.

Localization of Poincar\'e symmetry leads to Poincar\'e gauge theory of
gravity, which contains GR as a special case. Here, in contrast to GR,
at each point of spacetime there exists a whole class of local inertial
frames, mutually related by Lorentz transformations. Using this
freedom, allowed by the principle of equivalence, one can naturally
introduce not only {\it energy-momentum\/}, but also {\it spin\/} of
matter fields into gravitational dynamics.

We begin our exposition by presenting the basic principles of
Poincar\'e gauge theory (PGT). Then, an analysis of the geometric
interpretation of PGT leads us to conclude that spacetime has the
structure of Riemann--Cartan geometry, possessing both the {\it
curvature\/} and the {\it torsion\/}. Finally, we study in more details
the teleparallel limit of PGT, which represents a viable alternative
gravitational theory for macroscopic matter.

\section{Poincar\'e gauge theory}

We shall now analyze the process of transition from global to local
Poincar\'e symmetry, and find its relation to the gravitational
interaction [1--6]. Other spacetime symmetries (de Sitter, Weyl,
etc.) can by treated in an analogous manner \cite{7,6}.

\subsection*{Global Poincar\'e symmetry}

\prg{Minkowski spacetime.} In physical processes at low energies, the
gravitational field does not have a significant role, since the
gravitational interaction is extremely weak. In the absence of gravity,
the spacetime is described by the special relativity (SR), and its
mathematical structure corresponds to Minkowski space $M_4$. The
physical observer in spacetime makes use of some {\it reference
frame\/}, endowed with coordinates $x^\m$ ($\m=0,1,2,3$). An {\it
inertial observer\/} in $M_4$ can always choose global inertial
coordinates, such that the infinitesimal interval takes the simple form
$ds^2=\eta_{\m\n}dx^\m dx^\n$, where $\eta_{\m\n}=(1,-1,-1,-1)$ are
components of the metric in the inertial frame.

Coordinate transformations $x\to x'$ which do not change the form of
the metric define the {\it isometry group\/} of a given space. The
isometry group of $M_4$ is the group of global (rigid) Poincar\'e
transformations, the infinitesimal form of which is given by
\be
x'^\m=x^\m+\xi^\m(x)\, ,\qquad
\xi^\m=\om^\m{_\n}x^\n+\ve^\m \, ,                          \lab{1.1}
\ee
where $\om^{\m\n}=-\om^{\n\m}$ and $\ve^\m$ are ten constant
parameters (Lorentz rotations and translations).

\prg{Matter fields.} In order to define matter fields in spacetime
(scalars, spinors, etc.), it is useful to introduce the concept of
tangent space. The set of all tangent vectors at point $P$ in $M_4$
defines the tangent space $T_P$. Since the geometric structure of $M_4$
is pretty simple, the structure of $T_P$ actually coincides with that
of $M_4$. The choice of basis in the tangent space (frame) is not
unique. A coordinate frame ($C$ frame) is determined by a set of four
vectors $\mb{e}_\m$, tangent to the coordinate lines $x^\m$. In $M_4$
we can also introduce a local Lorentz frame ($L$ frame), represented by
a set of four orthonormal tangent vectors $\mb{e}_i(x)$ (vierbein or
tetrad): $\,\mb{e}_i\cdot\mb{e}_j=\eta_{ij}$.\footnote{ Here, the Latin
indices $(i,j,...)$ refer to local $L$ frames, while the Greek indices
$(\m,\n,...)$ refer to $C$ frames. Later, when we come to a geometric
interpretation, this distinction will become geometrically more
important.} To each $L$ frame $\mb{e}_i$ we can associate the related
(local) inertial coordinates $x^i$. If the coordinates $x^\m$ are
globally inertial, one can always choose the tetrad in such a way that
it coincides with the $C$ frame, $\mb{e}_i=\d_i^\m\mb{e}_\m$.

A matter field $\phi(x)$ in spacetime is always referred to an $L$
frame; in general, it is a multicomponent object which can be
represented as a vector-column. The action of global Poincar\'e
transformations in $T_P$ transforms each $L$ frame into another $L$
frame, inducing an appropriate transformation of the field $\phi(x)$:
$$
x'^i=x^i+\om^i{_j}x^j+\ve^i\, , \qquad
\phi'(x')=\left(1+\fr{1}{2}\om^{ij}\S_{ij}\right)\phi(x)\, .
$$
Here, $\S_{ij}$ is the spin matrix related to the multicomponent
structure of $\phi(x)$. Equivalently, we can write
\be
\d_0\phi= \left(\fr{1}{2}\om^{ij}M_{ij}+\ve^k P_k\right)\phi
        \equiv\cP\phi\, ,                                   \lab{1.2}
\ee
where $\d_0\phi(x)=\phi'(x)-\phi(x)$ is the change of form of
$\phi(x)$, and
$M_{ij}\equiv x_i\pd_j-x_j\pd_i+\S_{ij}$, $P_k\equiv-\pd_k$,
are the generators of global Poincar\'e transformations in the
space of fields.

Since form variation and differentiation are commuting operations,
we easily derive from \eq{1.2} the transformation law for
$\pd_k\phi$:
\be
\d_0\pd_k\phi=\cP\pd_k\phi+\om_k{^i}\pd_i\phi
             \equiv\cP_k{^m}\pd_m\phi\, .                   \lab{1.3}
\ee

\prg{Global Poincar\'e invariance.} Dynamical content of basic physical
interactions is successfully described by Lagrangian field theory.
Dynamical variables in this theory are fields $\phi(x)$, and dynamics
is determined by a function $\cL(\phi,\pd\phi)$, called the Lagrangian.
Equations of motion are given as the Euler--Lagrange equations for the
action integral $I=\int d^4 x\cL$.

Invariance of a theory under spacetime transformations can be expressed
in terms of some conditions on the Lagrangian, which are different from
those characterizing Yang--Mills theories. To see that, consider an
action integral defined over a spacetime region $\Om$,
$I(\Om)=\int_{\Om}d^4x \cL(\phi,\pd_k\phi;x)$, where we allow for the
possibility that $\cL$ may depend explicitly on $x$. The action
integral is invariant under spacetime transformations $x'=x+\xi(x)$ if
\cite{1}
\be
\D\cL \equiv\d_0\cL+\xi^\m\pd_\m\cL+(\pd_\m\xi^\m)\cL=0\, , \lab{1.4}
\ee
where
$\d_0\cL =(\pd\cL/\pd\phi)\d_0\phi+
          (\pd\cL/\pd\phi_{,k})\d_0\phi_{,k}$.
The Lagrangian $\cL$ satisfying the invariance condition \eq{1.4} is
usually called an invariant density. Here, we wish to make two
comments:
(a) the above result is based on the assumption $\d_0\eta_{\m\n}=0$;
(b) the condition \eq{1.4} can be relaxed by demanding a weaker
condition $\D\cL=\pd_\m\L^\m$; in that case the action changes by a
surface term, but the equations of motion remain invariant.

If we now substitute the Poincar\'e expressions for $\xi^\m$ and
$\d_0\phi$ in \eq{1.4}, the vanishing of the coefficients multiplying
$\om^{\m\n}$ and $\xi^\m$ implies two sets of identities: the first
identity is the condition of Lorentz invariance, while the second one,
related to translational invariance, is equivalent to the absence of
any explicit $x$ dependence in $\cL$, as one could have expected.

Assuming the equations of motion to hold, the invariance condition
\eq{1.4} leads to the differential conservation laws for Noether
currents --- the energy-momentum and angular momentum tensors. Spatial
integrals of the null components of the currents define the related
charges. The usual conservation in time of these charges does not hold
automatically, but only if the related flux integrals through the
boundary of the three-space vanish.

\subsection*{Localization of Poincar\'e symmetry}

Suppose now that we have a theory described by the matter Lagrangian
$\cL_M=\cL_M(\phi,\pd_k\phi)$, which is invariant under global
Poincar\'e transformations. If we now generalize Poincar\'e
transformations by replacing ten {\it constant\/} group parameters with
some {\it functions\/} of spacetime points, $\om^{ij}\to\om^{ij}(x)$,
$\ve^k\to\ve^k(x)$, the invariance condition \eq{1.4} is violated for
two reasons:
\bitem
\item[\bull] the old transformation rule \eq{1.3} of $\pd_k\phi$ is
changed into
\bea
\d_0\pd_k\phi &=&\cP\pd_k\phi -(\pd_k\xi^\n)\pd_\n\phi
             + \fr{1}{2}(\pd_k\om^{ij})\S_{ij}\phi \nn\\
   &=&\cP_k{^m}\pd_m\phi+(\pd\ve,\pd\om)\mbox{-terms}\, ;   \lab{1.5}
\eea
\item[\bull] the term $\pd_\m\xi^\m$ in \eq{1.4} does not vanish,
in contrast to the old relation $\pd_\m\xi^\m=\om^\m{_\m}=0$.
\eitem
The violation of local invariance can be compensated by certain
modifications of the original theory, whereupon the resulting theory
becomes locally invariant.

\prg{Covariant derivative.} In the first step of our compensation
procedure, we wish to eliminate non-invariance stemming from the
change of the transformation rule of $\pd_k\phi$. This can be
accomplished by introducing a new Lagrangian
\bsubeq
\be
\cL_M'=\cL_M(\phi ,\nabla_k\phi ) \, ,                     \lab{1.6a}
\ee
where $\nabla_k\phi$ is the {\it covariant derivative\/} of $\phi$,
which transforms according to the ``old rule" \eq{1.3}:
\be
\d_0\nabla_k\phi=\cP \nabla_k\phi +\om_k{^i}\nabla_i\phi\,.\lab{1.6b}
\ee
\esubeq
The new Lagrangian satisfies the condition
$\d\cL'_M\equiv\d_0\cL'_M+\xi\cdot\pd\cL'_M=0$.

The construction of $\nabla_k\phi$ is realized in two steps:
\bsubeq\lab{1.7}
\bea
i)&&\qquad \nabla_\m\phi =(\pd_\m + A_\m)\phi\, ,\qquad
           A_\m\equiv \fr{1}{2}A^{ij}{_\m}\S_{ij}\, ,    \lab{1.7a}\\
ii)&&\qquad \nabla_k\phi=\d_k^\m\nabla_\m\phi-A^\m{_k}\nabla_\m\phi
            \equiv h_k{^\m}\nabla_\m\phi\, .             \lab{1.7b}
\eea
\esubeq
The transformation rule of the $\om$-covariant derivative
$\nabla_\m\phi$,
\be
\d_0\nabla_\m\phi=\cP\nabla_\m\phi-(\pd_\m\xi^\n)\nabla_\n\phi\, ,
                                                            \lab{1.8}
\ee
is chosen so as to eliminate the term $\pd_k\om^{ij}$ appearing in
\eq{1.5}; it leads to a definite transformation rule for the Lorentz
compensating field $A^{ij}{_\m}$, given in \eq{1.9a}. The complete
$\nabla_k\phi$ is defined by adding a new field,
$h_k{^\m}=\d^\m_k-A^\m{_k}$, with the transformation properties defined
by equation \eq{1.6b}. It is useful to introduce another field
$b^s{_\n}$, the inverse of $h_k{^\m}$: $b^k{_\m}h_i{^\m}=\d^k_i$,
$b^k{_\m}h_k{^\n} = \d_\m^\n$. The transformation laws for the
compensating fields $A^{ij}{_\m}$ and $b^k{_\m}$ read:
\bsubeq\lab{1.9}
\bea
&&\d_0A^{ij}{_\m}= -\nabla_\m\om^{ij}
  -(\pd_\m\xi^\l)A^{ij}{_\l}-\xi^\l\pd_\l A^{ij}{_\m}\, ,\lab{1.9a}\\
&&\d_0 b^k{_\m}=\om^k{_s}b^s{_\m}-(\pd_\m\xi^\l)b^k{_\l}
              -\xi^\l\pd_\l b^k{_\m}\, ,                 \lab{1.9b}
\eea
\esubeq
with $\nabla_\m\om^{ij}=\pd_\m\om^{ij}
                 +A^i{}_{s\m}\om^{sj}+A_s{^j}{_\m}\om^{is}$.

In order to facilitate geometric interpretation of the local
transformations, it is convenient to generalize our previous convention
concerning the use of Latin and Greek indices. According to the
transformation rules \eq{1.9}, the use of indices in $A^{ij}{_\m}$ and
$b^k{_\m}$ follows the following convention: the fields transform as
local Lorentz tensors with respect to Latin indices, and as world
(coordinate) tensors with respect to Greek indices; the term
$-\pd_\m\om^{ij}$ shows that $A^{ij}{_\m}$ is not a true tensor but a
potential. One can also check that local Lorentz tensors can be
transformed into world tensors and vice versa, by the multiplication
with $h_k{^\m}$ or $b^k{_\m}$.

\prg{Matter field Lagrangian.} Up to now we have found the Lagrangian
$\cL'_M$, such that $\d\cL'_M=0$. In the second step of restoring local
invariance of the theory, we have to take care of the fact that
$\pd_\m\xi^\m\ne 0$. This can be done by introducing $\tcL_M=\L\cL'_M$,
where $\L$ is a suitable function of the new fields. The invariance
condition \eq{1.4} for $\tcL_M$ holds if $\d_0\L+\pd_\m(\xi^\m\L)=0$.
Using the known transformation properties of the fields, one finds a
simple solution for $\L$: $\L=\det (b^k{_\m})\equiv b$.

Thus, the final form of the modified Lagrangian for matter fields
is
\be
\tcL_M = b\cL_M(\phi ,\nabla_k\phi ) \, .                  \lab{1.10}
\ee
It is obtained from the original Lagrangian $\cL_M(\phi,\pd_k\phi)$ by
the minimal coupling prescription: a) $\pd_k\phi\to \nabla_k\phi$ and
b) $\cL_M\to b\cL_M$. The Lagrangian $\tcL_M$ satisfies the invariance
condition \eq{1.4} by construction, hence it is an invariant density.

The above construction is in general valid for massive matter fields.
In electrodynamics, however, one {\it can not\/} apply the prescription
$\pd\to\nabla$ without violating the internal $U(1)$ gauge symmetry!
Hence, in order to retain the internal gauge symmetry, one should keep
the original field strength unchanged, $F_{\m\n}=\pd_\m A_\m-\pd_\n
A_\m$. Although the minimal coupling prescription is thereby abandoned,
the procedure is compatible with {\it both\/} internal gauge symmetry
and local Poincar\'e covariance \cite{8}. Consequently, the
gravitational coupling to the electromagnetic field in PGT remains the
{\it same\/} as in GR.

\prg{Complete Lagrangian.} We succeeded to modify the original matter
Lagrangian by introducing gauge potentials, so that the invariance
condition \eq{1.4} remains true also for local Poincar\'e
transformations. In order to construct a Lagrangian for the new fields
$b^k{_\m}$ and $A^{ij}{_\m}$, we shall first introduce the
corresponding {\it field strengths\/}. The commutator of two covariant
derivatives has the form
$$
[\nabla_k,\nabla_l]\phi=\fr{1}{2}F^{ij}{}_{kl}\S_{ij}\phi
                         -F^s{_{kl}}\nabla_s\phi\, .
$$
Here, $F^{ij}{}_{kl}=F^{ij}{}_{\m\n}h_k{^\m}h_l{^\n}$,
$F^i{}_{kl}=F^i{}_{\m\n}h_k{^\m}h_l{^\n}$, and
\bea
&&F^{ij}{}_{\m\n}\equiv\pd_\m A^{ij}{_\n}-\pd_\n A^{ij}{_\m}
  +A^i{}_{s\m}A^{sj}{_\n}-A^i{}_{s\n}A^{sj}{}_\m\, ,\nn\\
&&F^i{_{\m\n}}\equiv\nabla_\m b^i{_\n}-\nabla_\n b^i{_\m}\, .
                                                           \lab{1.11}
\eea
The quantities $F^{ij}{_{\m\n}}$ and $F^i{_{\m\n}}$ are called
the Lorentz and translation field strengths, respectively.
They transform as tensors, in conformity with their index structure.

Jacoby identities for the commutators of covariant derivatives imply
the following {\it Bianchi identities\/} for the field strengths:
\bea
(\hbox{1st})&&\qquad\ve^{\r\m\l\n}\nabla_\m F^s{_{\l\n}}
  =\ve^{\r\m\l\n}F^s{_{k\l\n}}b^k{_\m}\, ,\qquad\quad \nn\\
(\hbox{2nd})&&\qquad\ve^{\r\l\m\n}\nabla_\l F^{ij}{_{\m\n}}=0\, .
                                                           \lab{1.12}
\eea

The free Lagrangian must be an invariant density depending only on the
field strengths, while the complete Lagrangian of matter and gauge
fields has the form
\be
\tcL=b\cL_F(F^{ij}{_{kl,}}F^i{_{kl}})
           +b\cL_M(\phi,\nabla_k\phi)\, .                  \lab{1.13}
\ee

\prg{Generalized conservation laws.} The invariance of the Lagrangian
in a gauge theory for an internal symmetry leads, after using the
equations of motion, to covariantly generalized differential
conservation laws. The same thing happens also in PGT. We restrict our
discussion to the matter Lagrangian $\tcL_M$, and introduce {\it
dynamical\/} energy-momentum and spin currents for matter fields:
$$
\t^\m{_k}= -{\d\tcL_M\over\d b^k{_\m}}\, ,\qquad
\s^\m{}_{ij}=-{\d\tcL_M\over\d A^{ij}{_\m}}\, .
$$
Assuming that matter field equations are satisfied, one can show
that local Poincar\'e invariance leads to generalized conservation laws
of the dynamical currents \cite{6}:
\bea
&&b^k{_\m}\nabla_\n\t^\n{_k}=\t^\n{}_{k}F^k{}_{\m\n}+
   \fr{1}{2}\s^\n{}_{ij}F^{ij}{}_{\m\n}\, ,\nn\\
&&\nabla_\m\s^\m{}_{ij}=\t_{ij}-\t_{ji}\, .                \lab{1.14}
\eea
Similar analysis can be applied to the complete Lagrangian \eq{1.13}.

\section{Geometric structure of spacetime}\setcounter{equation}{0}

In order to facilitate a proper understanding of the geometric content
of PGT, we introduce here some basic concepts of differential geometry
\cite{3,4,6,9}.

\subsection*{Riemann--Cartan geometry}

\prg{Manifolds.} Spacetime is often described as a ``four-dimensional
continuum". In SR, it has the structure of Minkowski space $M_4$. In
the presence of gravity spacetime can be divided into ``small, flat
pieces"  in which SR holds (on the basis of the principle of
equivalence), and these pieces are ``sewn together" smoothly. Although
spacetime looks locally like $M_4$, it may have quite different global
properties. Mathematical description of such four-dimensional
continuum is given by the concept of a differentiable manifold.

To be more rigorous, one should start with the natural concept of {\it
topological space\/}, which allows a precise  formulation of the idea
of continuity. A topological space $X$ is given the structure of a {\it
manifold\/} by introducing local coordinates on $X$. The compatibility
of different local coordinate systems promotes a manifold into a {\it
differentiable manifold\/}, in which one can easily introduce and study
mappings which are both continuous and differentiable.

\prg{Tensors.} Thus, we assume that spacetime has the structure of a
differentiable manifold $X_4$. We believe that the laws of physics can
be expressed as relations between geometric objects, such as
vectors, tensors, etc..

In order to define {\it tangent vectors\/} in terms of the internal
structure of the manifold, one should abandon the idea of the
``displacement" of a point. The most acceptable approach is to define
tangent vectors as directional derivatives, without any reference to
embedding. Directional derivatives represent an abstract realization of
the usual geometric notion of tangent vectors.

The set of all tangent vectors at $P$ defines the tangent space $T_P$.
The set of vectors tangent to the coordinate lines $x^\m$ defines the
coordinate basis $\mb{e}_\m=\pd_\m$ in $T_P$. An arbitrary vector
\mb{v} in $T_P$ can be represented in the form $\mb{v}= v^\m\mb{e}_\m$,
where $v^\m$ are components of \mb{v} in the basis $\mb{e}_\m$. Under
the change of local coordinates $x\mapsto x'$, both $\mb{e}_\m$ and
$v^\m$ change the form,
$$
\mb{e}'_\m=\frac{\pd x^\n}{\pd x'^\m}\mb{e}_\n\, ,\qquad
v'^\m=\frac{\pd x'^\m}{\pd x^\n}v^\n\, ,
$$
but \mb{v} itself remains invariant. The second equation is known as
the vector transformation law. Vectors $\mb{v}=(v^\m)$ are usually
called contravariant vectors.

Following the usual ideas of linear algebra, we can associate a dual
vector space $T^*_P$ with each tangent space $T_P$ of $X$. Consider
linear mappings from $T_P$ to $R$, defined by
$\mb{w}^*:~\mb{v}\mapsto\mb{w}^*(\mb{v})\in R$. If a set of these
mappings is equipped with the usual operations of addition and scalar
multiplication, we obtain the dual vector space $T^*_P$. Vectors
$\mb{w}^*$ in $T^*_P$ are called {\it dual vectors\/}, covariant
vectors (covectors) or differential forms. Given the basis
$\mb{e}_\m$ in $T_P$, one can construct its dual basis $\th^\m$ in
$T^*_P$ by demanding $\th^\m(\mb{v})=v^\m$, or equivalently,
$\th^\m(\mb{e}_\n)=\d^\m_\n$. Each dual vector $\mb{w}^*$ can be
represented in the form $\mb{w}^*=w^*_\m\th^\m$. A change of local
coordinates induces the following change in $\th^\m$ and $w^*_\m$:
$$
\mb{\th}'^\m=\frac{\pd x'^\m}{\pd x^\n}\mb{\th}^\n\, ,\qquad
w^{*\prime}_\m=\frac{\pd x^\n}{\pd x'^\m}w^*_\n\, .
$$
To simplify the notation, one usually omits the sign $^*$ for dual
vectors.

The concept of a dual vector as a linear mapping from $T_P$ to $R$, can
be naturally extended to the concept of {\it tensor\/} as a multilinear
mapping. Thus, a tensor \mb{\om} of type (0,2) is a bilinear mapping
which maps a pair of vectors (\mb{u},\mb{v}) into a real number
\mb{\om}(\mb{u},\mb{v}). Using the dual basis $\th^\m\otimes\th^\n$, we
can represent \mb{\om} as $\mb{\om}=\om_{\m\n}\th^\m\otimes\th^\n$, so
that $\mb{\om}(\mb{u},\mb{v})=\om_{\m\n}u^\m v^\n$. Similarly, a tensor
\mb{\a} of type (1,1) maps a pair $(\mb{u},\mb{w}^*)$ into a real
number $\mb{\a}(\mb{u},\mb{w}^*)=\a_\m{^\n}u^\m w_\n$. After these
examples, it is not difficult to define the general tensor \mb{t} of
type $(p,q)$. Its components transform as the product of $p$ vectors
and $q$ dual vectors.

A {\it tensor field\/} on $X$ is a mapping $x\mapsto \mb{t}(x)$ that
associates a tensor $\mb{t}(x)$ to each point $x$ in $X$.

Totally antisymmetric tensor fields of type $(0,p)$ are particularly
important objects, called {\it differential $p$-forms\/} (forms of
degree $p$). A 1-form \mb{\a} is a dual vector,
$\mb{\a}=\a_\m\mb{\th}^\m$. A 2-form \mb{\b} in the basis
$\mb{\th}^\m\otimes\mb{\th}^\n$ is given as
$\mb{\b}=\fr{1}{2}\b_{\m\n}\mb{\th}^\m\wedge\mb{\th}^\n$, where
$\mb{\th}^\m\wedge\mb{\th}^\n\equiv
\mb{\th}^\m\otimes\mb{\th}^\n-\mb{\th}^\n\otimes\mb{\th}^\m$, and so
on. In the space of smooth $p$-forms one can introduce the {\it
exterior derivative\/} as a differential operator \mb{d} which maps a
$p$-form \mb{\a} into a $(p+1)$-form \mb{d\a}.

{\it Tensor densities\/} are objects similar to tensors; they can be
defined on orientable manifolds.

\prg{Parallel transport.}  On $X_4$ one can define differentiable
mappings, tensors, and various algebraic operations with tensors at a
given point (addition, multiplication, contraction). However, comparing
tensors at different points requires some additional structure on
$X_4$: the law of parallel transport. Consider, for instance, a smooth
vector field $\mb{A}$ on $X$, such that $\mb{A}_x$ lies in the tangent
space $T_x$, and $\mb{A}_y$ is in $T_y$. In order to compare $\mb{A}_x$
with $\mb{A}_y$, it is necessary first to ``transport" $\mb{A}_x$ from
$T_x$ to $T_y$, and then to compare the resulting object
$\mb{A}_\pt\equiv(\mb{A}_\pt)_y$ with $\mb{A}_y$. This ``transport"
procedure generalizes the concept of parallel transport in flat space
and bears the same name. The vector $\mb{A}_\pt$ is in general
different from $\mb{A}_y$. If the point $y$ is infinitesimally close to
$x$, $y=x+dx$, then the components of $\mb{A}_y$ with respect to the
coordinate basis at $y$ have the form $A^\m_y=A^\m(x+dx)$, while those
of $\mb{A}_\pt$ are defined by the rule (Figure 1)
\be
A_\pt^\m=A^\m(x)+\d A^\m(x)\, ,\qquad
   \d A^\m=-\G^\m_{\l\r}A^\l dx^\r \, ,                     \lab{2.1}
\ee
where the infinitesimal change $\d A^\m$ is bilinear in $A^\l$ and
$dx^\r$. The set of 64 components $\G^\m_{\l\r}$ defines a linear (or
affine) {\it connection\/} \mb{\G} on $X_4$, in the coordinate basis.
An $X_4$ equipped with \bG\ is called {\it linearly connected space\/},
$\bL_4=(X_4,\bG)$.
\begin{figure}[htb]
\begin{center}
\includegraphics[height=3.5cm]{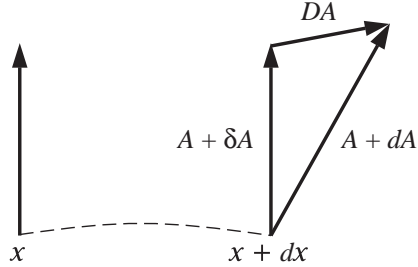}
\end{center}
\vspace*{-.5cm}
\caption{Parallel transport of vectors from $T_x$ to $T_{x+dx}$}
\end{figure}

Linear connection is equivalently defined by the {\it covariant
derivative\/} \mb{D}. Computing, for instance, the difference
$\mb{A}_y-\mb{A}_\pt$ we find
\bea
DA^\m&\equiv& A^\m_y-A^\m_\pt=dA^\m-\d A^\m  \nn\\
  &=&(\pd_\r A^\m +\G^\m_{\l\r}A^\l)dx^\r
  \equiv D_\r(\G)A^\m dx^\r\, .                             \lab{2.2}
\eea
Covariant derivative of a dual vector is defined by demanding
$\d(A^\m B_\m)=0$. Covariant derivative of an arbitrary
tensor field $\mb{t}(p,q)$ is defined a) as a mapping
$\mb{t}(p,q)\mapsto D\mb{t}(p,q+1)$, which is b) linear, satisfies
the Leibnitz rule, $Df=df$ if $f$ is a scalar, and commutes with
contraction.

The linear connection is not a tensor, but its antisymmetric part defines
a tensor called the {\it torsion tensor\/}:
\be
T^{\m}{_{\l\r}}=\G^\m_{\r\l}-\G^\m_{\l\r}\, .               \lab{2.3}
\ee

Parallel transport is a path dependent concept. If we parallel
transport a vector around an infinitesimal closed path, the result is
proportional to the {\it Riemann curvature tensor\/}:
\be
R^\m{_{\n\l\r}}=\pd_\l\G_{\n\r}^\m
      +\G_{\s\l}^\m\G_{\n\r}^\s-(\l\lra\r)\, .              \lab{2.4}
\ee

\prg{Metric compatible connection.} On $X_4$ one can define {\it metric
tensor} \mb{g} as a symmetric, nondegenerate tensor field of type
$(0,2)$. After that we can introduce the scalar product of two tangent
vectors, $\mb{u}\cdot\mb{v}=\mb{g}(\mb{u},\mb{v})$, and calculate
lengths of curves, angles between vectors, etc. The differentiable
manifold $X_4$ equipped with linear connection and metric becomes {\it
linearly connected metric space\/}
$(\bL_4,\mb{g})\equiv(X_4,\bG,\mb{g})$.

Generally, linear connection and metric are {\it independent\/}
geometric objects. In order to preserve lengths and angles under
parallel transport in $(\bL_4,\mb{g})$, one can impose the {\it metricity
condition\/}
\be
-Q_{\m\n\l}\equiv D_{\m}g_{\n\l}=\pd_{\m}g_{\n\l}
     -\G_{\n\m}^{\r}g_{\r\l}-\G_{\l\m}^{\r}g_{\n\r}=0\, ,   \lab{2.5}
\ee
which relates $\bG$ and \mb{g}. The requirement of vanishing
nonmetricity \mb{Q} establishes local Minkowskian structure
on $X_4$, and defines a metric compatible linear connection:
\be
\G^\m_{\l\n}=\Chr{\m}{\l\n}+K^\m{}_{\l\n}\, ,\qquad
K^\m{}_{\l\n}\equiv-\fr{1}{2}(T^\m{}_{\l\n}
             -T_\n{}^\m{}_\l+T_{\l\n}{}^\m)\, ,             \lab{2.6}
\ee
where $\Chr{\m}{\l\n}$ is the Christoffel connection and
$K$ the contortion.

A space $(\bL_4,\mb{g})$ with the most general metric compatible linear
connection \bG\ is called {\it Riemann--Cartan\/} space $U_4$. If the
torsion vanishes, a $U_4$ becomes a {\it Riemannian\/} space $V_4$ of GR;
if, alternatively, the curvature vanishes, a $U_4$ becomes
Weitzenb\"ock's {\it teleparallel\/} space $T_4$. Finally, the condition
$\,R^\m{_{\n\l\r}}=0\,$ transforms a $V_4$ into a Minkowski space $M_4$,
and $\,T^\m{_{\l\r}}=0\,$ transforms a $T_4$ into an $M_4$ (Figure 2).
\begin{figure}[htb]
\begin{center}
\includegraphics[height=4.2cm]{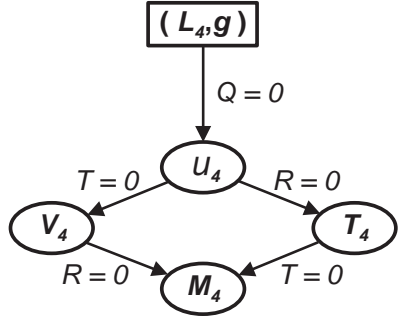}
\end{center}
\vspace*{-.5cm}
\caption{Classification of spaces satisfying the metricity condition}
\end{figure}

\prg{Spin connection.} Linear connection and metric are geometric
objects independent of the choice of frame. Their components are
defined with respect to a frame and are, clearly, frame-dependent. The
choice of frame in $T_P$ is not unique; $C$ frames $\mb{e}_\m$ and $L$
frames $\mb{e}_i$ are of particular practical importance.\footnote{The
existence of $L$ frames is closely related to the principle of
equivalence.} Every tangent vector \mb{u} can be expressed in both
frames: $\mb{u}=u^\m\mb{e}_\m=u^i\mb{e}_i$. In particular,
$\mb{e}_i=e_i{^\m}\mb{e}_\m$, $\mb{e}_\m=e^i{_\m}\mb{e}_i$, and
accordingly, $u^i=e^i{_\m}u^\m$, $u^\m=e_i{^\m}u^i$.
The scalar product of two tangent vectors can be written in two
equivalent forms: $\mb{u}\cdot\mb{v}=g_{\m\n}u^\m v^\n
=\eta_{ij} u^i v^j$, where
$$
\eta_{ij}=\mb{e}_i\cdot\mb{e}_j=g_{\m\n}e_i{^\m}e_j{^\n}\, ,\qquad
g_{\m\n}=\mb{e}_\m\cdot\mb{e}_\n=\eta_{ij}e^i{_\m}e^j{_\n}\, .
$$

The parallel transport of a tangent vector $\mb{u}\in T_x$, represented
in the form $\mb{u}=u^i\mb{e}_i$, is defined by the parallel
transport rule
$$
\d u^i =-\om^i{}_{j\m}u^j dx^\m\, ,
$$
where $\om^{ij}{_\m}$ is the so-called {\it spin connection\/}, with
64 components. Parallel transport of $v_i$ is determined by requiring
$\d(u^i v_i)=0$: $\d v_i =\om^j{}_{i\m}v_j dx^\m$.
An equivalent definition of the parallel transport may be given in
terms of the $\om$-covariant derivative:
\be
D(\om)u^i=\bigl(\pd_\m u^i+\om^i{}_{j\m}u^j\bigr)dx^\m
         \equiv D_\m(\om)u^i dx^\m \, ,                     \lab{2.7}
\ee
and similarly for $v_i$.

The existence of $L$ frames at each point of $X_4$ implies the
existence of the Lorentz metric $\eta_{ij}$ at each point of $X_4$.
Demanding the tensor field $\eta_{ij}$ to be invariant under the
parallel transport, implies that the connection is antisymmetric
in its Latin indices:
$$
\d\eta_{ij}=\bigl(\om^s{}_{i\m}\eta_{sj}
            +\om^s{}_{j\m}\eta_{si}\bigr)dx^\m
           =\bigl(\om_{ji\m}+\om_{ij\m}\bigr)dx^\m=0\, .
$$
Since \mb{\eta} is a constant tensor, its covariant derivative
vanishes:
\be
D_\m(\om)\eta_{ij}=0\, .                                    \lab{2.8}
\ee

\prg{Relation between \mb{\om} and \mb{\G}.} The parallel transport is
a unique geometric operation, independent of the choice of frame, hence
$$
u^i+\d u^i=(u^\m+\d u^\m)e^i{_\m}(x+dx)\, .
$$
From this property, we obtain the relation between \mb{\om} and
\mb{\G\,}, called the {\it tetrad postulate\/}:
\be
D_\m(\om+\G)e^i{_\n}\equiv D_\m(\om)e^i{_\n}
                           -\G^\r_{\n\m}e^i{_\r}=0\, ,      \lab{2.9}
\ee
where $D_\m(\om)e^i{_\m}=\pd_\m e^i{_\n}+\om^i{}_{j\m}e^j{_\n}$.
The operator $D(\om+\G)$ can be formally understood as a ``total"
covariant derivative. Using the above equations we easily derive the
metricity condition:
$$
D_\m(\G)g_{\m\n}=D_\m(\om+\G)g_{\m\n}
      =D_\m(\om+\G)(\eta_{ij}e^i{_\m}e^j{_\n})=0\, .
$$

The $\om$-covariant derivative can be generalized to a quantity
$\phi$ belonging to an arbitrary representation of the Lorentz group:
\be
D_\m(\om)\phi=\bigl(\pd_\m+\om_\m\bigr)\phi\, ,\qquad
  \om_\m\equiv\fr{1}{2}\om^{ij}{_\m}\S_{ij}\, ,             \lab{2.10}
\ee
where $\S$ is the related spin matrix.

It is interesting to note that if we find $\G=\G(\om)$ from equation
\eq{2.9} and substitute the result into the expressions \eq{2.3} and
\eq{2.4} for the torsion and the curvature, we obtain
\bsubeq\lab{2.11}
\bea
&&T^\r{}_{\m\n}e^i{_\r}=D_\m(\om)e^i{_\n}-D_\n(\om)e^i{_\n}
    \equiv T^i{}_{\m\n}(\om)\, ,                          \lab{2.11a}\\
&&e^i{_\l}e^{j\r}R^\l{}_{\r\m\n}
    =\pd_\m\om^{ij}{_\n}+\om^i{}_{k\m}\om^{kj}{_\n}-(\m\lra\n)
    \equiv R^{ij}{}_{\m\n}(\om)\, .                       \lab{2.11b}
\eea
\esubeq
Equation \eq{2.11a} can be formally solved for the connection $\om$:
\bea
&&\om_{ij\m}=\D_{ij\m}+K_{ij\m}\, ,\nn\\
&&\D_{ij\m}\equiv
  \fr{1}{2}(c_{ijm}-c_{mij}+c_{jmi})e^m{_\m}\, ,            \lab{2.12}
\eea
where $c^i{}_{\m\n}=\pd_\m e^i{_\n}-\pd_\n e^i{_\m}$ is called the
object of anholonomity, and $K$ is the contortion.

\subsection*{Geometric interpretation of PGT}

The final result of the analysis of PGT is the construction of the
invariant Lagrangian \eq{1.12}. It is achieved by introducing new
fields  $A^{ij}{_\m}$ and $b^i{_\m}$ (or $h_k{^\n}$), which are used to
construct the covariant derivative $\nabla_k=h_k{^\n}\nabla_\n$ and the
field strengths $F^{ij}{_{\m\n}}$ and $F^i{_{\m\n}}$. This theory can
be thought of as a field theory in Minkowski space. However, geometric
analogies are so strong, that it would be unnatural to ignore them.

\def\rl{\rule[-1.2ex]{0pt}{3ex}}

\begin{center}
\doublerulesep 1.5pt
\begin{tabular}{|l|l|}
\multicolumn{2}{l}{\rl Table 1. Relation between PGT (left) and
                    Riemann--Cartan geometry}\\  \hline\hline
\rule[-1pt]{0pt}{15pt}
      \rl $A^{ij}{_\m}$ Lorentz gauge field
     &$\om^{ij}{_\m}$ spin connection\\               
\rule{0pt}{12pt}
      \rl $\nabla_\m(A)$ covariant derivative
     &$D_\m(\om)$ $\om$-covariant derivative\\        \hline
\rule[-1pt]{0pt}{15pt}
      \rl $b^i{_\m}~~(h_k{^\n})$ transl. gauge field
     &$e^i{_\m}~~(e_k{^\n})$ tetrad coframe (frame)\\ \hline\hline
\rule{0pt}{12pt}
     \rl $\G^\r_{\m\n}$ not defined
    &$\G^\r_{\m\n}$ connection\\                      
\rule[-1pt]{0pt}{15pt}
     \rl{\it tetrad postulate!\/} $\Ra$ $\G^\r_{\m\n}$ defined
    &$D_\m(\om)e^i{_\n}=\G^\r_{\n\m}e^i{_\r}$ tetrad postulate\\
                                                      \hline
\rule[-1pt]{0pt}{15pt}
     \rl $\nabla_\m\eta_{ij}=0$ metricity condition
    &$D_\m(\G)g_{\n\l}=0$ metricity condition\\      \hline\hline
\rule[-1pt]{0pt}{15pt}
     \rl $F^i{}_{\m\n}=\nabla_\m b^i{_\n}-\nabla_\n b^i{_\m}$
    &$T^i{}_{\m\n}=(\G^\r_{\n\m}-\G^\r_{\m\n})b^i{_\r}
                          = T^i{}_{\m\n}(\om)$\\      
\rule[-1pt]{0pt}{15pt}
     \rl $F^{ij}{}_{\m\n}=F^{ij}{}_{\m\n}(A)$
    &$R^{ij}{}_{\m\n}=R^{ij}{}_{\m\n}(\om)$\\        \hline\hline
\end{tabular}
\end{center}\smallskip

$\a)$ The Lorentz gauge field $A^{ij}{_\m}$ can be identified with the
spin connection $\om^{ij}{_\m}$, as follows from its transformation law
\eq{1.9a}. Equivalently, $\nabla_\m(A)$ can be identified with the
geometric covariant derivative $D_\m(\om)$. This follows from the
definition of $\nabla_\m\phi$, which implies that a) the quantity
$\nabla_\m\phi$ has one additional dual index, as compared to $\phi$;
b) it acts linearly, obeys Leibniz rule, commutes with contraction, and
$\nabla_\m f=\pd_\m f$ if $f$ is a scalar function.

$\b)$ The field $b^i{_\m}$ can be identified with $e^i{_\m}$,
on the basis of its transformation law \eq{1.9b}. It ensures the
possibility to transform local Lorentz and coordinate indices into each
other.

$\g)$ Local Lorentz symmetry of PGT implies the metricity condition
\eq{2.8}. After adopting the tetrad postulate \eq{2.9}, whereby one
introduces the connection $\G^\r_{\n\m}$ in PGT, the metricity
condition \eq{2.8} becomes equivalent to \eq{2.5}.

$\d)$ It follows from equations \eq{2.11} that the translation field
strength $F^i{_{\m\n}}$ is nothing but the torsion $T^i{}_{\m\n}$,
while the Lorentz field strength $F^{ij}{}_{\m\n}$ represents the
curvature $R^{ij}{}_{\m\n}$.
\bitem
\item[\bull] Consequently, PGT has the geometric structure of
Riemann--Cartan space $U_4$.
\eitem

Although PGT has a well defined geometric interpretation, its gauge
structure differs from what we have in ``standard" gauge theories
(Appendix A).

\subsection*{The principle of equivalence in PGT}

The principle of equivalence (PE) is a dynamical principle, which
severely restricts the form of the gravitational interaction. It states
that $i)$ the effect of gravity on matter is locally equivalent to the
effect of a non-inertial reference frame in special relativity
(SR).\footnote{The PE does not allow, for instance, the $\phi R$
coupling of the scalar matter.}

To clarify the dynamical content of the PE, let us consider an inertial
frame in $M_4$, in which (massive) matter field $\phi$ is described by
a Lagrangian $\cL_M(\phi,\pd_i\phi)$. When we go over to a non-inertial
frame, $\cL_M$ transforms into $b\cL_M(\phi,\nabla_i\phi)$, with
$\nabla_i=e_i{^\m}(\pd_\m+\om_\m)$. The pseudo-gravitational field,
equivalent to the non-inertial reference frame, is contained in $b$ and
$\nabla_i$. This field can be eliminated on the whole spacetime by
simply going back to the global inertial frame, while for real
gravitational fields this is not true --- they can be eliminated only
locally, as we shall see. For this reason, in the last step of
introducing a real gravitational field, Einstein replaced $M_4$ with a
Riemann space $V_4$. Although this is a correct choice, we shall see
that Einstein could have chosen also a Riemann--Cartan space $U_4$.

Let us now recall another formulation of the PE:~ $ii)$ the effect of
gravity on matter can be locally eliminated by a suitable choice of
reference frame, whereupon matter behaves as in SR. More precisely,
\bitem
\item[\bull] at any point $P$ in spacetime one can choose an
orthonormal reference frame $\mb{e}_i$, such that:
a)~$\om^{ij}{_\m}=0$,~~b)~$e_i{^\m}=\d_i^\m$,~~at $P$.
\eitem
We shall see that this statement is correct not only in GR ($V_4$), but
also in PGT ($U_4$) \cite{10,11}.

Gravitational theory in Riemann space $V_4$ possesses certain features
which do not follow necessarily from the PE. Namely, the form of
Riemannian connection shows that relative orientation of the
orthonormal frame $\mb{e}_i(x+dx)$ with respect to $\mb{e}_i(x)$
(parallel transported to $x+dx$) is completely fixed by the metric. The
change of this orientation is described by Lorentz transformations,
which do not produce any gravitational effect; therefore, there is no
reason to prevent any {\it additional\/} Lorentz rotation of local
frames. If we want to realize this freedom, the spin connection should
contain an extra part, independent of the metric:
$\om^i{}_{j\m}\equiv\D^i{}_{j\m}+K^i{}_{j\m}$. Interpreted in this way,
the PE becomes nicely incorporated into Riemann--Cartan geometry, as
shown bellow.

Let $\mb{e}_i$ be a basis for $T_P$ in spacetime. For each $i$, one can
define an auto-parallel $C_i$ through $P$, with tangent vector
$\mb{e}_i$. By parallel transporting $\mb{e}_i$, one can define a
vector field $\mb{e}_i$ along $C_i$, in some neighborhood $\cO_i$.
Taking a suitable restriction of the intersection of all $\cO_i$, we
can find a neighborhood $\cO$ of $P$ in which the auto-parallels $C_i$
do not intersect. The set of vector fields $\mb{e}_i$ can be extended
to form a {\it parallel\/} frame on $\cO$. Hence, the connection
coefficients at $P$, defined with respect to this particular frame,
vanish: $\G^i_{jk}(P)=0$. This result makes no use of any metric
property, and holds for an arbitrary linearly connected manifold
\cite{11}. In $U_4$, the parallel frame on $\cO$ can be made {\it
orthonormal\/}, $\mb{e}_i\cdot\mb{e}_j=\eta_{ij}$, which gives an
attractive physical content to the PE. Using the formula $\G^i_{jk}=
e_k{^\r}\om^i{}_{j\r}$, we conclude that $\om^{ij}{_\m}(P)=0$.

At each point $P$ in $\cO$, one can introduce local inertial
coordinates, defined by $dx^i=e^i{_\m}dx^\m$. Let us now change the
coordinates $x^\m$, $x^\m\to y^\m$, so that $y^\m$ coincide with $x^i$
at $P$: $dy^\m =\d^\m_i dx^i$. This coordinate transformation ensures
$e_i{^\m}(P)=\d_i^\m$, without changing $\om^{ij}{_\m}(P)=0$.
\bitem
\item[\bull] The existence of torsion does not violate the PE.
\eitem

The PE fits naturally into a $U_4$ geometry of spacetime. In
particular, it holds in $V_4$, and also in $T_4$. In more general
geometries, where the symmetry of the tangent space is higher than the
Poincar\'e group, the usual form of the PE is violated, and local
physics is different from SR \cite{7,6}.

\section{The teleparallel theory}\setcounter{equation}{0}

Dynamics of the gravitational field in PGT is determined by the form of
the gravitational Lagrangian $\cL_G$. If we demand that the equations
of motion are at most of second order in field derivatives, $\cL_G$ can
be at most quadratic in torsion and curvature. A lot of different
invariants makes the general structure of $\cL_G$ rather complicated:
$\cL_G\sim R+T^2+R^2+\l$, with eleven $(1+3+6+1)$ constant parameters
\cite{5}.

The simple action
$$
I_{\rm EC}=\int d^4x b(-aR+\cL_M)\, ,\qquad a=\frac{1}{16\p G}\, ,
$$
defines the so-called Einstein--Cartan (EC) theory, a direct
generalization of GR to Riemann--Cartan spacetime \cite{1,2}. The EC
theory incorporates both mass and spin of matter as sources of the
gravitational field, and represent a description of gravity which is
{\it microscopically\/} more satisfying than GR. Indeed, in current
theories of fundamental interactions matter is described by matter
fields, with their spins, symmetries and conserved currents; at this
level, there is no space for the conventional GR, with matter
consisting of point particles, fluids and light rays. On the other
hand, spin effects are negligible for macroscopic matter, so that the
empirical predictions of the EC theory are, for all practical purposes,
the same as in GR. A simple but accurate  way to depict this situation
is to name GR ``the best available {\it alternative\/} gravitational
theory", the best theory being the EC theory itself \cite{12}. Since
the structure of the EC theory is pretty well known, we turn our
attention to the teleparallel theory as ``the next best" alternative
\cite{12}.

General geometric arena of PGT, the Riemann--Cartan space $U_4$, may be
{\it a priori\/} restricted by imposing certain conditions on the
curvature and the torsion. Thus, Einstein's GR is defined in Riemann
space $V_4$, obtained from $U_4$ by the requirement of vanishing
torsion. Another interesting limit of PGT is Weitzenb\"ock or  {\it
teleparallel\/} geometry $T_4$, defined by the requirement
\be
R^{ij}{}_{\m\n}(A)=0 \, .                                   \lab{3.1}
\ee
The vanishing of curvature means that parallel transport is path
independent (if some topological restrictions are adopted), hence we
have an absolute parallelism. The teleparallel geometry is, in a
sense, complementary to Riemannian: curvature vanishes, and torsion
remains to characterize the parallel transport.

The physical interpretation of the teleparallel geometry is based on
the fact that there is a one-parameter family of teleparallel
Lagrangians which is {\it empirically\/} equivalent to GR
\cite{13,14,15}.

\prg{Lagrangian.} In the framework of the teleparallel geometry $T_4$,
gravitational field is described by the tetrad $b^k{_\m}$ and Lorentz
connection $A^{ij}{_\m}$, subject to the condition of vanishing
curvature. We shall consider here the gravitational dynamics determined
by a class of Lagrangians quadratic in the torsion:
\bea
&&\tcL=b\cL_T +\l_{ij}{}^{\m\n}R^{ij}{}_{\m\n}+\tcL_M\, ,\nn\\
&&\cL_T=a\bigl(AT_{ijk}T^{ijk}+BT_{ijk}T^{jik}+CT_{k}T^{k}\bigr)
       \equiv \b_{ijk}(T)T^{ijk} \, ,                       \lab{3.2}
\eea
where $\l_{ij}{}^{\m\n}$ are Lagrange multipliers introduced to ensure
the teleparallelism condition \eq{3.1} in the variational formalism,
and $\b_{ijk}=a\bigl(AT_{ijk}+BT_{[jik]}+C\eta_{i[j}T_{k]}\bigr)$.

The parameters $A,B,C$ in the Lagrangian should be determined on
physical gro\-unds, so as to obtain a consistent theory which could
describe all the known gravitational experiments. If we require that
the theory \eq{3.2} gives the same results as GR in the li\-ne\-ar,
weak-field approximation, we can restrict our considerations to the
one-parameter family of Lagrangians, defined by the conditions
\cite{13,14,15}
\par {$i)$} $\,2A+B+C=0\, ,\quad C=-1$.

\ni This family represents a viable gravitational theory for
macroscopic matter (scalar and electromagnetic fields), empirically
indistinguishable from GR. Von der Heyde and Hehl have given certain
theoretical arguments in favor of the choice $B=0$ \cite{4}.
There is, however, another, particularly interesting choice determined
by the requirement
\par {$ii)$} $\,2A-B=0$.

\ni In the gravitational sector, this choice leads effectively to the
Einstein--Hilbert Lagrangian of GR, $\cL_{EH}=-abR(\D)$, with
Riemannian connection $A=\D$. To see that, we substitute the expression
\eq{1.12} for the spin connection, $A=\D+K$, into the definition of the
scalar curvature tensor $R(A)$, and obtain the geometric identity
\be
abR(A)=abR(\D)+\cL_T^{\parallel}-2a\pd_\r(bT^\r)\, ,          \lab{3.3}
\ee
where $\cL_T^{\parallel}$ is the torsion Lagrangian \eq{3.2} with
\be
A=\fr{1}{4}\, ,\qquad B=\fr{1}{2}\, ,\qquad C=-1  \, .      \lab{3.4}
\ee
The conditions $i)$ and $ii)$ given above coincide with \eq{3.4}. In
the teleparallel spacetime, where $R(A)=0$, the identity \eq{3.3}
implies the relation $\cL_T^{\parallel}=\cL_{EH}$ + divergence; that is
why the teleparallel theory \eq{3.2} with $\cL_T=\cL_T^{\parallel}$ is
called the teleparallel formulation of GR (\tgr). It is equivalent to
GR for scalar and electromagnetic matter (see Lecture 1), but the other
matter fields have different couplings in $T_4$ and $V_4$.

\prg{Field equations.} By varying the Lagrangian \eq{3.2} with
respect to $b^i{_\m}, A^{ij}{_\m}$ and $\l_{ij}{}^{\m\n}$, we obtain
the gravitational field equations \cite{16}:
\bsubeq\lab{3.5}
\bea
&&4\nabla_\r\bigl(b\b_i{^{\m\r}}\bigr)-4b\b^{nm\m}T_{nmi}
     +h_i{^\m}b\cL_T =\t^\m{_i}\, ,                       \lab{3.5a}\\
&&4\nabla_\r\l_{ij}{}^{\m\r}
     -8b\b_{[ij]}{^\m}=\s^\m{}_{ij} \, ,                  \lab{3.5b}\\
&&R^{ij}{}_{\m\n}=0 \, .                                  \lab{3.5c}
\eea
\esubeq

The third field equation defines the teleparallel geometry in PGT. The
first field equation is a dynamical equation for $b^k{_\m}$.
The symmetric part of this equation plays the role analogous to
Einstein's equation in GR, while the antisymmetric part implies
\bsubeq\lab{3.6}
\be
4\nabla_\m(b\b_{[ij]}{^\m})\approx\t_{[ji]}\, .            \lab{3.6a}
\ee
In \tgr, the left hand side vanishes, so that $\t_{[ij]}$ must also
vanish. Since this is not true for Dirac field, it follows that the
description of Dirac matter in \tgr\ is {\it not consistent\/}. In the
one-parameter teleparallel theory the left hand side is proportional to
the axial torsion, and we do not have any problem. By taking the
covariant divergence of \eq{3.5b}, one obtains the consistency
condition
\be
-8\nabla_\m\bigl( b\b_{[ij]}{^\m}\bigr)
                     \approx\nabla_\m\s^\m{}_{ij}\, .      \lab{3.6b}
\ee
\esubeq
This condition is satisfied as a consequence of \eq{3.6a} and the
second identity in \eq{1.14}. Thus, the only role of \eq{3.5b} is to
determine the Lagrange multipliers $\l_{ij}{}^{\m\n}$. Taking into
account equation \eq{3.6b}, one concludes that {\it the number of
independent equations \eq{3.5b} is $24-6=18$\/}. It is clear that these
equations cannot determine 36 multipliers $\l_{ij}{}^{\m\n}$ in a
unique way. As we shall see, non-uniqueness of $\l_{ij}{}^{\m\n}$ is
related to an extra gauge freedom in the theory.

\prg{The \mb{\lambda} symmetry.} The gravitational Lagrangian \eq{3.2}
is, by construction, invariant under the local Poincar\'e
transformations. In addition, it is also invariant, up to a
four-divergence, under the transformations \cite{16}
\bsubeq
\be
\d_0\l_{ij}{}^{\m\n}=\nabla_\r\ve_{ij}{}^{\m\n\r} \, ,     \lab{3.7a}
\ee
where the gauge parameter $\ve_{ij}{}^{\m\n\r}=-\ve_{ji}{}^{\m\n\r}$ is
completely antisymmetric in its upper indices, and has $6\times 4=24$
components. The invariance is easily verified by using the second
Bianchi identity $\ve^{\l\r\m\n}\nabla_\r R^{ij}{}_{\m\n}=0$. On the
other hand, the invariance of the field equation \eq{3.5b} follows
directly from $R^{ij}{}_{\m\n}=0$. The symmetry \eq{3.7a} will be
referred to as $\l$ symmetry.

It is useful to observe that the $\l$ transformations can be written in
the form
\be
\d_0\l_{ij}{}^{\a\b}=\nabla_0\ve_{ij}{}^{\a\b}
                     +\nabla_\g\ve_{ij}{}^{\a\b\g}\, ,\qquad
\d_0\l_{ij}{}^{0\b}=\nabla_\g\ve_{ij}{}^{\b\g}\, ,         \lab{3.7b}
\ee
\esubeq
where $\ve_{ij}{}^{\a\b}\equiv\ve_{ij}{}^{\a\b 0}$.
However, one can show by canonical methods (Appendix B) that
\bitem
\item[\bull] the only independent parameters of the $\l$ symmetry are
$\ve_{ij}{}^{\a\b}$.
\eitem
The six parameters $\ve_{ij}{}^{\a\b\g}$ are not independent of
$\ve_{ij}{}^{\a\b}$. Hence, they can be completely discarded, leaving
us with $24-6=18$ {\it independent gauge parameters\/}. They can be
used to fix 18 multipliers $\l_{ij}{}^{\m\n}$, whereupon the remaining
18 multipliers are determined by the independent field equations
\eq{3.5b} (at least locally).

The Poincar\'e and $\l$ gauge symmetries are always present ({\it
sure\/} symmetries), independently of the values of parameters $A,B$
and $C$ in the teleparallel theory \eq{3.2}. Specific models, such as
\tgr, may have {\it extra\/} gauge symmetries,  which are present only
for some critical values of the parameters. The gauge structure of the
one-parameter teleparallel theory is problematic \cite{17}.

\prg{OT frames.} Teleparallel theories in $U_4$ are based on the
condition of vanishing curvature. Let us choose an arbitrary
tetrad at point $P$ of spacetime. Then, by parallel transporting this
tetrad to all other points, we generate the tetrad field on spacetime
manifold. If the manifold is paralellizable (which is a strong
topological assumption), the vanishing of curvature implies that the
parallel transport is path independent, so the resulting tetrad field is
globally well defined. In such an {\it orthonormal\/} and {\it
teleparallel\/} (OT) frame, the connection coefficients vanish:
\be
A^{ij}{_\m}=0 \, .                                          \lab{3.8}
\ee
The above construction is not unique --- it defines a class of OT frames,
related to each other by global Lorentz transformations. In an OT frame,
the covariant derivative reduces to the partial derivative, and the
torsion takes the simple form:
$T^i{}_{\m\n}=\pd_\m b^i{_\n}-\pd_\n b^i{_\m}$ (see e.g. \cite{18}).

Equation \eq{3.8} defines a particular solution of the condition
$R^{ij}{}_{\m\n}(A)=0$. Since a local Lorentz transformation of the
tetrad field induces a non-homogeneous change in the connection,
$$
e'{^i}{_\m}=\L^i{_k}e^k{_\m} \quad \Ra \quad
A'^{ij}{_\m}=\L^i{_m}\L^j{_n}A^{mn}{_\m}+\L^i{_m}\pd_\m\L^{jm}\, ,
$$
it follows that the general solution of $R^{ij}{}_{\m\n}(A)=0$ has the
form $A^{ij}{_\m}=\L^i{_m}\pd_\m \L^{jm}$. Thus, the choice \eq{3.8}
breaks local Lorentz invariance, and represents a {\it gauge fixing\/}
condition.

In the action \eq{3.2}, the condition of teleparallelism is ensured by
the Lagrange multiplier. The field equation \eq{3.5b} merely serves to
determine the multiplier, while the non-trivial dynamics is completely
contained in \eq{3.5a}. Hence, the teleparallel theory (on
parallelizable manifolds) may also be described by imposing the gauge
condition \eq{3.8} directly in the action. The resulting theory is
defined in terms of the tetrad field only, and may be thought of as the
gauge theory of translations. This formalism is often used in the
literature because of its technical simplicity, but the local
Lorentz-invariant formulation simplifies the canonical analysis of the
conservation laws.

\prg{Exact solutions.} It is interesting to see how some exact
solutions of the one-parameter theory can be obtained by a simple
analysis of the field equations \cite{19,20}. We start with the torsion
Lagrangian of the one-parameter theory, written in the form
\be
\cL_T=\cL_T^\parallel-\frac{a}{12}(2B-1)\cA_{ijk}\cA{}^{ijk}\, ,
\qquad  \cA_{ijk}\equiv T_{ijk}+T_{kij}+T_{jki}\, ,         \lab{3.9}
\ee
where $\cL_T^\parallel$ is the torsion Lagrangian of \tgr. Using the
geometric identity \eq{3.3}, it follows that the first field equation
has the form
\be
R_{ik}(\D)-\fr{1}{2}\eta_{ik}R(\D)+\cO(\cA)=\t_{ki}/2ab\, ,\lab{3.10}
\ee
where $\cO(\cA)$ are terms proportional to $\cA_{ijk}$. For $2B-1=0$
the above Lagrangian reduces to the GR form, and equations \eq{3.10}
coincides with Einstein's equations. More generally, for any field
configurations satisfying $\cA_{ijk}=0$, {\it the first field equation
has the same form as in GR\/}. The consistency of this equation
requires $\t_{ki}$ to be symmetric.

Taking into account that the second field equation serves only to
determine the Lagrange multipliers $\l_{ij}{}^{\m\n}$, we can use this
result to generate some solutions of the teleparallel theory, starting
from certain solutions of GR. Consider, for instance, a metric which
has diagonal form in some coordinate system:
\bsubeq
\be
ds^2=A(dx^0)^2-B_1(dx^1)^2-B_2(dx^2)^2-B_3(dx^3)^2\, .    \lab{3.11a}
\ee
Let us choose the tetrad components to be diagonal,
\be
b^0{}_0=\sqrt{A}\, ,\qquad b^a{_\a}=\d^a_\a\sqrt{B_\a}\, ,\lab{3.11b}
\ee
\esubeq
and fix the gauge $A^{ij}{}_\m=0$. Then, one easily proves that
$\cA_{ijk}=0$, and derives an important consequence:
\bitem
\item[\bull] If the diagonal metric \eq{3.11a} is a solution
of GR, the related tetrad \eq{3.11b} is a solution of the one-parameter
theory, in the gauge $A^{ij}{_\m}=0$ and with the same $\t_{ik}$.
\eitem
An important class of solutions of this type is the class of
spherically symmetric solutions.

All observational differences from GR are related to the effects
stemming from $\cA_{ijk}\ne 0$.

\prg{On the physical interpretation.} We have seen that the field
equations of \tgr\ are identical to those of GR for macroscopic matter
(scalar and electromagnetic fields), but the coupling of Dirac field is
not consistent. What happens in the one-parameter theory? The related
argument about Dirac matter coupling does not hold any more: the
antisymmetric part of the first field equation \eq{3.10} shows that
$\t_{[ij]}$ is proportional to the axial torsion $\cA_{klm}$ contained
in $\cO(\cA)$. Thus, it seems that one should abandon \tgr\ and use the
one-parameter theory in order to consistently describe Dirac matter.
However, serious arguments given in Refs. \cite{17} strongly indicate
that the gauge structure of the one-parameter theory (the initial value
problem and the canonical formulation) is problematic. Unless the
problem is solved in a satisfactory way, one should remain skeptical
about the idea of treating this theory as a fundamental approach to
gravity.

The situation just described led some authors to interpret the
teleparallel theory only as an {\it effective\/} macroscopic theory of
gravity \cite{14}. If we accept this point of view, we can investigate
experimental predictions of the theory by using test particles/fields
of any type (scalars, spinors, etc). Possible empirical differences
between GR and the one-parameter teleparallel theory can be tested by
measuring non-trivial axial torsion effects \cite{19,21}.

\section*{Concluding remarks}

We conclude the exposition with a short summary.

1) PGT is based on the global Poincar\'e symmetry,  a well
established symmetry in particle physics, and incorporates both mass
and spin as sources of the gravitational field.

2) The geometric interpretation of PGT leads to Riemann--Cartan
geometry of spacetime, in which both curvature and torsion are used to
characterize the gravitational phenomena. Riemann--Cartan geometry is
compatible with the principle of equivalence.

3) The EC version of PGT is microscopically more satisfying then GR,
while its macroscopic predictions are, for all practical purposes, the
same as in GR.

4) In the teleparallel limit of PGT, curvature vanishes and torsion
remains to characterize both the geometry of spacetime and the
gravitational dynamics. The general one-parameter theory, including
\tgr\ as a special case, is empirically equivalent to GR. In spite of
that, the existing consistency problems make it difficult to accept the
teleparallel theory as a fundamental theory of gravity.

\subsection*{Acknowledgements}

I would like to thank M. Vasili\'c, F. W. Hehl and J. Garecki for
useful discussions.

\appendix
\subsection*{Appendix A: On the gauge structure of PGT}
\setcounter{section}{1}\setcounter{equation}{0}

It is an intriguing fact that PGT does not have the structure of an
``ordinary" gauge theory \cite{22}. To clarify this point, we start
from the Poincar\'e generators $P_a,M_{ab}$ and define the gauge
potential as $A_\m=e^a{_\m}P_a+\fr{1}{2}\om^{ab}{_\m}M_{ab}$. The
infinitesimal gauge transformation
$$
\d_0 A_\m=-\tilde\nabla_\m\l\equiv-\pd_\m \l-[A_\m,\l]\, ,
$$
where $\l=\l^a P_a+\fr{1}{2}\l^{ab}M_{ab}$, has the following component
content:
\bea
\hbox{Translations:}&&\qquad
 \d_0 e^a{_\m}=-\nabla_\m\l^a\, ,\qquad \d_0\om^{ab}{_\m}=0\, ,\nn\\
   \hbox{Rotations:}&&\qquad
       \d_0 e^a{_\m}=\l^a{_b}e^b{_\m}\, ,\qquad~
       \d_0\om^{ab}{_\m}=-\nabla_\m\l^{ab}\, , \nn
\eea
where $\nabla=\nabla(\om)$ is the covariant derivative with respect
to the spin connection $\om$. The resulting gauge transformations are
clearly {\it different\/} from those obtained in PGT.

Although the tetrad field and the spin connection carry a
representation of the Poincar\'e group, the EC action in four
dimensions, $I_{EC}=\fr{1}{4}\int d^4x
\ve^{\m\n\l\r}\ve_{abcd}e^c{_\l}e^d{_\r}R^{ab}{}_{\m\n}$, is {\it
not\/} invariant under the translational part of the Poincar\'e group,
$$
\d_T I_{EC}=\fr{1}{4}\int d^4x
   \ve^{\m\n\l\r}\ve_{abcd}\l^c T^d{}_{\l\r}R^{ab}{}_{\m\n}\ne 0\, ,
$$
but it remains invariant under Lorentz rotations and diffeomorphisms.
The situation is different in 3d, where gravity {\it can\/} be
represented as a ``true" gauge theory \cite{23}.

\subsection*{Appendix B: Canonical generator of the \mb{\l} symmetry}
\setcounter{section}{2}\setcounter{equation}{0}

The canonical analysis of a gauge theory is the best way to explain its
gauge structure. We apply this approach to examine the $\l$ symmetry in
the teleparallel theory \cite{16}.

The basic phase space dynamical variables of the teleparallel theory
\eq{3.2} are $(b^i{}_\m,A^{ij}{}_\m,\l_{ij}{}^{\m\n})$ and the
corresponding momenta $(\p_i{}^\m,\p_{ij}{}^\m,\p^{ij}{}_{\m\n})$.
Going through the standard Dirac type analysis, one can find all the
constraints and the total Hamiltonian. Then, starting from the primary
first class constraint $\pi^{ij}{}_{\a\b}$, one can apply the general
canonical procedure and show that the canonical gauge generator acting
on the Lagrange multipliers $\l_{ij}{}^{\m\n}$ has the form \bsubeq
\bea
&&G_A[\ve]= \frac{1}{4}\int d^3x
     \left[\dot\ve_{ij}{}^{\a\b}\p^{ij}{}_{\a\b}
    +\ve_{ij}{}^{\a\b}S^{ij}{}_{\a\b}\right]\, , \nn\\
&&S^{ij}{}_{\a\b}\equiv -4\left(R^{ij}{}_{\a\b}
    -\fr{1}{2}\nabla_{[\a}\pi^{ij}{}_{0\b]} \right)
               + 2 A^{[i}{}_{k0}\,\p^{j]k}{}_{\a\b}\, .    \lab{B.1a}
\eea
Using the rule $\d_0 X = \int d^3x'\{X,G'\}$, we apply the generator
\eq{B.1a} to the $\l$ field, and find
\be
\d_0^A\l_{ij}{}^{0\a}=\nabla_{\b}\,\ve_{ij}{}^{\a\b}\, ,\qquad
\d_0^A\l_{ij}{}^{\a\b}=\nabla_0\,\ve_{ij}{}^{\a\b}\, ,     \lab{B.1b}
\ee
\esubeq
as the only non-trivial transformations. Surprisingly, this result does
not agree with the form of the $\l$ symmetry \eq{3.7b}, which contains
an additional piece, $\nabla_\g\ve_{ij}{}^{\a\b\g}$, in the expression
for $\d_0\l_{ij}{}^{\a\b}$. Since there are no other primary first
class constraints that could produce the transformation of
$\l_{ij}{}^{\a\b}$, the canonical origin of the additional term seems
somewhat puzzling.

The solution of the problem is, however, quite simple: if we consider
only {\it independent\/} gauge transformations, this term is not
needed. To prove this statement, consider the following primary first
class constraint
$$
\Pi^{ij}{}_{\a\b\g}=\nabla_\a\p^{ij}{}_{\b\g}
     +\nabla_\g\p^{ij}{}_{\a\b}+\nabla_\b\p^{ij}{}_{\g\a}\, ,
$$
which is essentially a linear combination of $\p^{ij}{}_{\a\b}$. Hence,
the related gauge generator will not be truly independent of the
general expression \eq{B.1a}. The standard canonical construction yields
\bsubeq
\be
G_B[\ve]=-\frac{1}{4}\int d^3x\,\ve_{ij}{}^{\a\b\g}
              \nabla_{\a}\p^{ij}{}_{\b\g}\, ,              \lab{B.2a}
\ee
where the parameter $\ve_{ij}{}^{\a\b\g}$ is totally antisymmetric
with respect to its upper indices. The only non-trivial field
transformation produced by this generator is
\be
\d_0^B\l_{ij}{}^{\a\b}=\nabla_\g\ve_{ij}{}^{\a\b\g} \, ,   \lab{B.2b}
\ee
\esubeq
and it coincides with the missing term in equation \eq{B.1b}. Thus,
if we are interested only in the independent $\l$ transformations,
\bitem
\item[\bull] the six parameters $\ve_{ij}{}^{\a\b\g}$ in the $\l$
transformations \eq{3.7b} can be completely discarded.
\eitem
Although the generator $G_B$ is not truly independent of $G_A$, it is
convenient to define $G\equiv G_A+G_B$ as an overcomplete gauge
generator, since it automatically generates the covariant Lagrangian
form of the $\l$ symmetry.

\renewcommand{\refname}{{\large References}}

\end{document}